\documentclass{ws-fml}
\usepackage{multicol}
\begin{document}
\markboth{Richard Herrmann}{A fractal approach to  the dark silicon problem}

\title{A fractal approach to  the dark silicon problem: \\
a comparison of 3D computer architectures - \\
standard slices versus fractal Menger sponge geometry   }

\author{Richard Herrmann}

\address{gigaHedron\\
Berliner Ring 80, D-63303 Dreieich, Germany
\\
\email{herrmann@gigahedron.com}
}

\maketitle

\begin{history}
\received{April 7, 2014}
\revised{\today}
\end{history}

\begin{abstract}
The dark silicon problem, which limits the power-growth of future computer generations,  is interpreted as a heat energy transport problem when increasing the energy emitting  surface area within a given volume. A comparison of two 3D-configuration models, namely a standard slicing and a fractal surface generation within the
Menger sponge geometry is presented. It is shown, that for iteration orders $n>3$ the fractal model shows increasingly better thermal behavior. As a consequence cooling problems may be minimized by using a fractal architecture.  Therefore the Menger sponge geometry is a good example for fractal architectures applicable not only in computer science, but also e.g. in chemistry when building chemical reactors,  optimizing catalytic processes or in sensor construction technology building highly effective sensors for toxic gases or water analysis.        
\end{abstract}

\keywords{fractal architecture, Menger sponge, computer science, dark silicon, chemical reactors, catalytic reactions, sensor technology}

\begin{multicols}{2}
\section{Introduction}
According to Moore's law\cite{Moo65} the size of computer chips and the price per transistor is decreasing exponentially for more than 40 years now. But in the future this development will come to an end due to the problem of heat production on multi-core chips. 
Despite the fact, that many attempts have been made to reduce heat effects in chips, e.g. reducing the power consumption, the design of state of the art chips comes to its limits.

Nowadays, a single chip produces energy densities comparable with nuclear power plants which turns out to be the major obstacle in upscaling multicore CPUs\cite{Esm11} and
server-farms respectively\cite{Har11}. The phrase "dark silicon" drastically describes the consequences for actual multi-core CPUs: large areas on a chip may only be used for a limited time and have to be switched off for a cooling period of time, which is steadily increasing with chip size. 

One option, to overcome these difficulties, is the design of alternative chip architectures\cite{Tay12}.
A promising direction is to make a step from 2D to 3D design\cite{Hua06}.  In three dimensional space the control of heat generation within a given volume becomes an even more important task,  which until now is an open problem in 3D-chip design and a field of actual research\cite{Rah01}.

Let us state the problem with the following words: In 3D-chip design we have to find structures, which on one hand guarantee a maximum surface within a given volume, which means maximum computing power per volume and on the other hand a minimum of volume consumption,  such that a cooling medium may occupy a maximal portion of the given volume, which results in maximum heat transfer and cooling.  

To solve this problem, we suggest to consider a fractal approach, which is motivated by the observation, that similar problems have already been solved by evolutionary processes in nature:

One example is the epithelial tissue, the material lungs are made of, which combines the requirements of optimum gas diffusion with minimum volume consumption and reveals a fractal structure under the microscope\cite{Wei09}.  

To model such a fractal structure, in the next section we will consider the properties of the Menger sponge\cite{Men26}, which may be considered as the 3D-extension of the 2-dimensional Sierpinski carpet\cite{Sie16}, which itself already serves as a blue print for 2D architectures in technical systems, like building high quality antennas for mobile phones\cite{Pue98}. 

It will be demonstrated, that there exists a threshold iteration number $n=3$, above which a Menger sponge like architecture shows increasingly better thermal characteristics compared to a standard multi-core sliced ingot.

\section{The models}
We will consider two different models and investigate their thermal behavior when increasing the active surface area within a given unit cube $V$ with size 1. Its dimensions may be measured in $[m^3]$ to describe a server farm, in $[cm^3]$ to describe a multi-processor CPUs or in $[mm^3]$ to describe a 3D-storage unit. 

We consider the thermal active areas to be equivalent to the surfaces $S_\textrm{model}$ on a passive 3D-substrate, which generates the volume $V_\textrm{model}$ of the presented model.  

The first model configuration is a simple sliced ingot:

Introducing an iteration variable $n$, which is correlated with an increasing number of slices  of height $L$ \begin{equation}
L = \frac{1}{3^n}
\end{equation}
and volume $1 \times 1 \times L$,
we divide the unit-cube into $\rho$ different slices, 
\begin{equation}
\rho = \textrm{floor} (3^n/2)+1
\end{equation}
which are evenly spaced within the cube and distance $L$ apart from each other, such that a cooling medium, that fills the empty space, may be used for heat transport with a given velocity $v$.       
The volume $V_s$ occupied by the slices follows as
\begin{equation}
V_s = \rho L 
\end{equation}
\begin{figurehere}
\centerline{\psfig{file=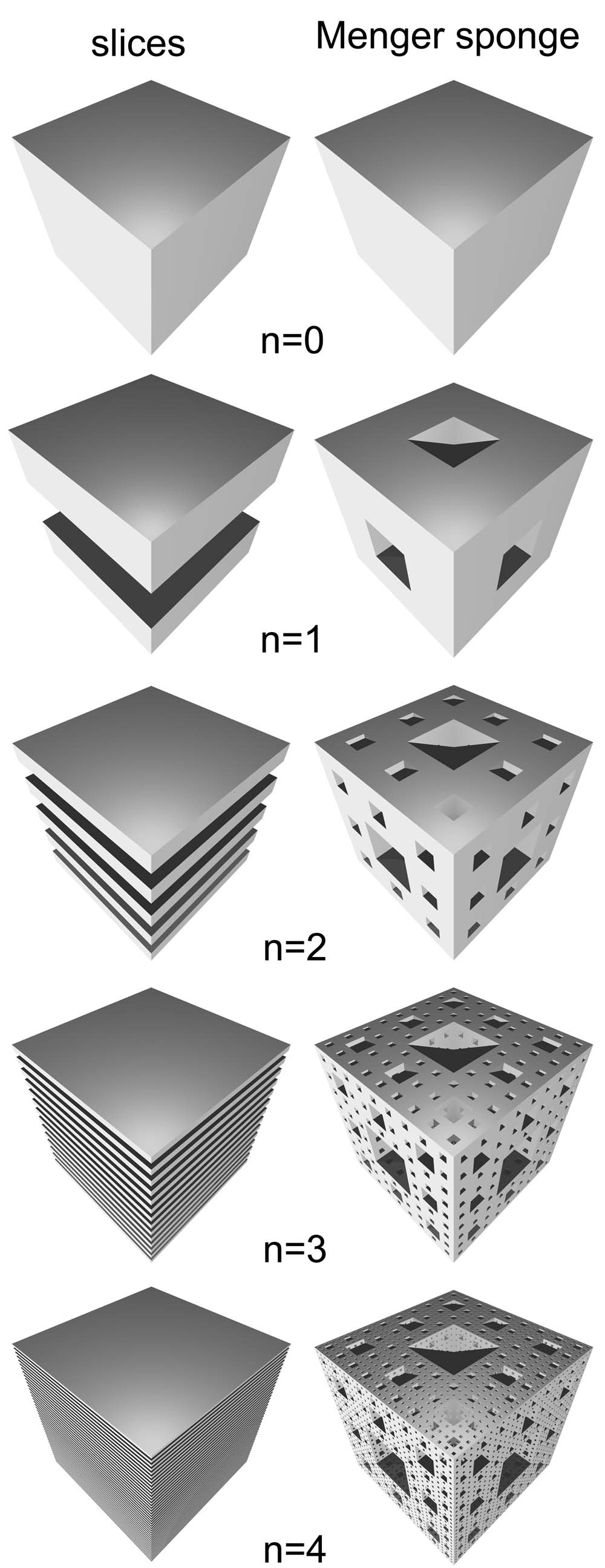,width=8.4cm}}
\caption{A sequence of 3D-configurations with increasing iteration value $n$ is shown. The left column illustrates the standard slice configuration,  the right columns illustrates the fractal Menger sponge model.}
\label{fig1}
\end{figurehere}
\noindent and the corresponding surface $S_s$ follows as
\begin{equation}
S_s = \rho ( 2 + 4 L) 
\end{equation}
In the left column of figure \ref{fig1} the resulting configurations for increasing $n$ are shown.

The second model configuration is the fractal Menger sponge. For a given iteration value $n$, the 
volume $V_M$ is given as
\begin{equation}
V_M = (\frac{20}{27})^n 
\end{equation}
 and the corresponding surface $S_M$ follows as
\begin{equation}
S_M = (\frac{1}{9})(\frac{20}{9})^{n-1} (40+80 (\frac{2}{5})^n) 
\end{equation}
In the right column of figure \ref{fig1} the resulting configurations for increasing $n$ are shown.

In order to compare the thermal properties of each configuration, we insert the objects into  a wraping  cube of size $V_{\textrm{tot}}$
\begin{equation}
V_{\textrm{tot}} =(1 + 2 L)^3 
\end{equation}
which is considered as the container for a cooling medium, which covers a volume $V_c$
\begin{equation}
V_c = V_{\textrm{tot}} -  V_\textrm{model}
\end{equation}
A measure of the efficiency $E_\textrm{model}$ of a given configuration is the ratio of available cooling volume per unit surface
\begin{eqnarray}
E_s &=& \frac{V_{\textrm{tot}} -  V_s}{S_s}\\
E_M &=& \frac{V_{\textrm{tot}} -  V_M}{S_M}
\end{eqnarray}
 where smaller values indicate a more and more problematic thermal configuration, because this value must be compensated by an increasing heat capacity  and/or an increasing exchange flow velocity $v$ of the cooling medium.     

\begin{table*}
\tbl{For increasing iteration value $n$, resulting number of slices $\rho$, characteristic length $L$, occupied volume and surface,  total volume $V_{\textrm{tot}}$ including boundary surfaces, efficiencies $E$ and ratios $R$ for the two different presented models are shown. In the last column the quality ratio $R_n$ for the two models is shown. For $R_n>1$ the fractal model is better. $(-p)$ is short hand notation for $\times 10^{-p}$, $p \in \mathbb{N}$. }
{\begin{tabular}{@{}lrlrrrrrlllll@{}}
\toprule
$n$ & $\rho$ & $L$    & $V_M$    & $V_s$     & $S_M$  & $S_s$   & $V_{\textrm{tot}}$ & $E_M$          &  $E_s$  & $R_E$ & $R_S$ & $R_n$   \\ \colrule
0    &   1      & 1      & 1.0000 &1.0000     &   6.0000 & 6.0000  &27.0000 & 4.3334 & 4.3334 & 1.0000& 1.0000& 1.0000\\
1    &   2      & 1/3   &  0.7407 &0.6667    &   8.0000  & 6.6667  & 4.6296 & 4.8611(-1) & 5.9444(-1) & 0.8177& 1.2000&0.9813\\
2    &   5      & 1/9   &  0.5487 &0.5556    &   13.0370  & 12.2222  & 1.8258 & 9.7959(-2) & 1.0393(-1) & 0.9426&1.0667&1.0054\\
3    & 14      & 1/27  &  0.4064 &0.5185   &24.7572  &30.0741 &1.2391 & 3.3633(-2) & 2.3910(-2) & 1.4037&0.8232&1.1555\\
4    & 41      & 1/81  &  0.3011 &0.5062   & 51.2702  &84.0247 &1.0759 & 1.5113(-2) & 6.7807(-3) & 2.2288&0.6102&1.3599\\
5    & 122    & 1/243 &  0.2230 &0.5021   & 110.604  &246.008 &1.0249 & 7.2500(-3) & 2.1253(-3) & 3.4113&0.4496&1.5337\\
6    & 365    & 1/729 &  0.1652 &0.5007   & 242.828  &732.003 &1.0083 & 3.4718(-3) & 6.9339(-4) & 5.0070&0.3317&1.6610\\
\botrule
\end{tabular}}\label{tab:smtab}
\end{table*}

\section{Discussion}
For increasing iteration values $n=0,...,6$ we have calculated the characteristic volumes, surfaces and efficiencies for the two different models. The results are listed in table 1. 

Introducing the ratio $R_n$
\begin{eqnarray}
R_n &=& R_E R_S \\
& =& \frac{E_M}{E_s}\frac{S_s}{S_E}
\end{eqnarray}
\begin{figurehere}
\centerline{\psfig{file=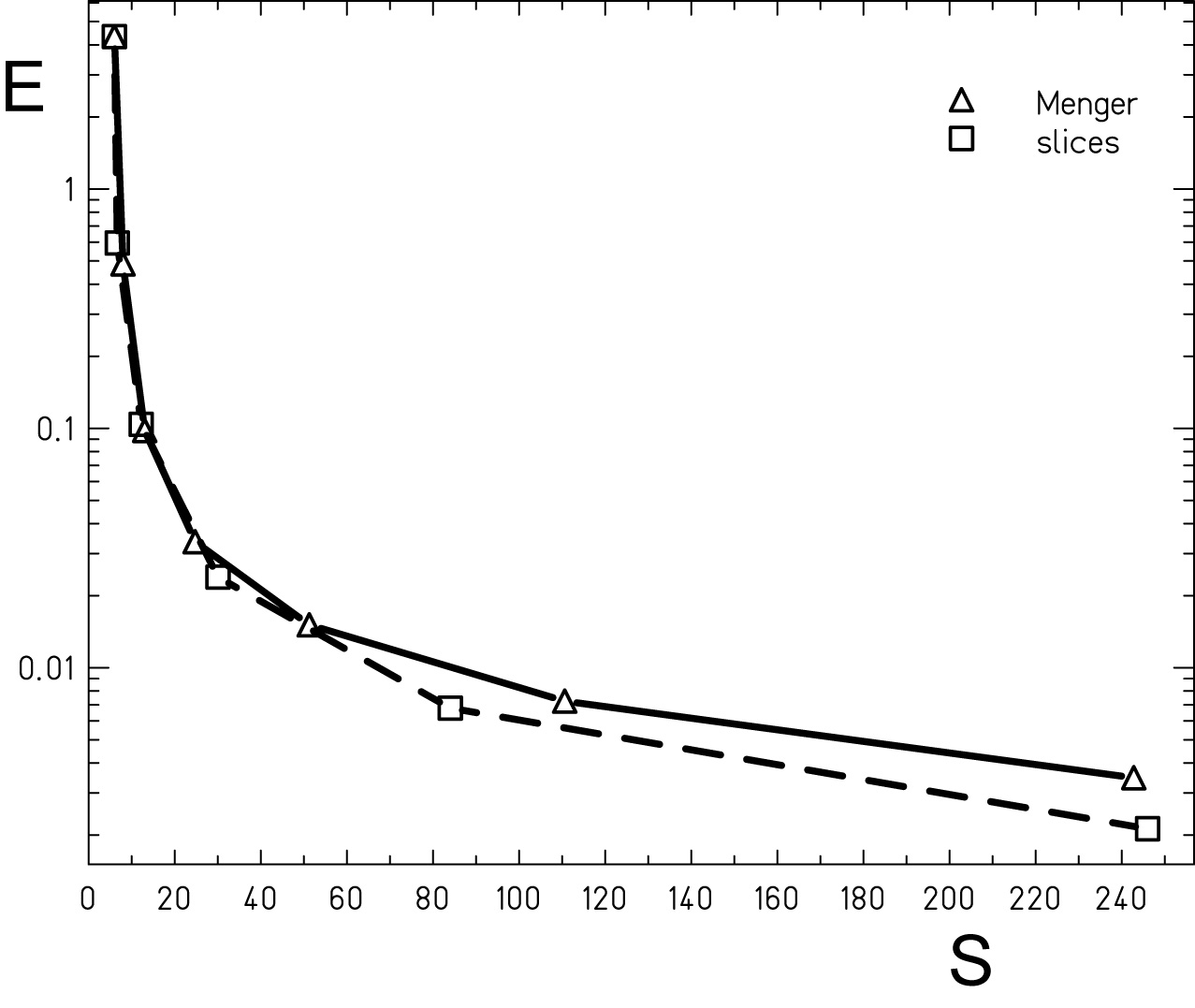,width=8.7cm}}
\caption{As a function of surface size $S$ the efficiency $E_\textrm{model}$ is plotted for Menger sponge (triangles)
and slice model (squares). For surface sizes $S > 50$ the Menger sponge geometry becomes more efficient.}   
\label{figMenger}
\end{figurehere}
\noindent which is a measure for the quality of the thermal properties, a value $R_n>1$ indicates, that the fractal Menger sponge configuration
shows a better behavior compared to the standard slices.

In order to compare the results directly, in figure \ref{figMenger} we have plotted the efficiency $E_\textrm{model}$ for a given surface size  $S$, the values are directly extracted from table 1. 

While for $S<50$ the standard slice model is favorable with respect to its thermal properties, the configuration based on the fractal Menger sponge model becomes the increasingly  better model for surface size $S>50$. 

Therefore we observe a topological phase transition with respect to the thermal properties for the presented geometries. 

The reason for this increasing quality is the fractal geometry of the Menger sponge, where the volume $V_M$ is reduced for 
increasing iteration value, while for the slice model the volume tends to $V_s \rightarrow 1/2$ and as a consequence, there is less space left for the cooling medium.

In addition, the slice geometry becomes more and more mechanically unstable, compared to the robust configuration of the fractal Menger sponge.  
 
\section{Conclusion} 
As a result of our schematic study we may conclude, that the fractal Menger sponge architecture may serve as an promising alternative to standard
3D geometries with a superior thermal behavior.     

The fractal approach, we have considered here,  fulfills both requirements simultaneously: a maximized surface area is implemented within a given volume and the ratio of active contact medium versus passive substrate is optimized. 

Therefore the Menger sponge geometry is a good example for fractal architectures not only applicable in computer science, but also e.g. in chemistry when building chemical reactors or optimizing catalytic processes. Sensor technology would benefit from highly efficient sensors for toxic gases or water analysis.  

Since the Menger sponge geometry is just one example for fractal architectures,  a still open question is the search for an optimum fractal geometry and may be a subject of future research. 
\nonumsection{Acknowledgments} \noindent 
We thank A. Friedrich for helpful discussions.

\end{multicols}
\end{document}